\documentclass[aps,pre,twocolumn,showpacs,floatfix,amsmath]{revtex4}
\usepackage{epsfig}
\usepackage[latin1]{inputenc}
\usepackage{bm} 

\begin{document}

\title{Binary threshold networks as a natural null model for biological networks}
\author{Matthias Rybarsch}
\email{rybarsch@itp.uni-bremen.de}
\author{Stefan Bornholdt}
\email{bornholdt@itp.uni-bremen.de}
\affiliation{Institute for Theoretical Physics, University of Bremen, D-28359 Bremen, Germany}  
\bibliographystyle{apsrev}
\begin{abstract}
Spin models of neural networks and genetic networks are considered elegant as they are accessible to statistical mechanics tools for spin glasses and magnetic systems. However, the conventional choice of variables in spin systems may cause problems in some models when parameter choices are unrealistic from a biological perspective. Obviously, this may limit the role of a model as a template model for biological systems. Perhaps less obviously, also ensembles of random networks are affected and may exhibit different critical properties. We consider here a prototypical network model that is biologically plausible in its local mechanisms. We study a discrete dynamical network with two characteristic properties: Nodes with binary states 0 and 1, and a modified threshold function with $\Theta_0(0)=0$. We explore the critical properties of random networks of such nodes and find a critical connectivity $K_c=2.0$ with activity vanishing at the critical point. Finally, we observe that the present model allows a more natural implementation of recent models of budding yeast and fission yeast cell-cycle control networks. 
\end{abstract}

\maketitle

\section{Introduction}
There has been a revived interest recently in discrete dynamical networks of nodes with binary states, driven by two active fields of research: modeling of molecular information processing networks (e.g., genetic networks or protein networks) \cite{Bornholdt:2008}, as well as modeling of adaptive networks \cite{Gross:2008}. These network models with binary states and discrete update time steps are reminiscent of artificial neural networks as studied in the statistical mechanics community about two decades ago. 

An early motivation of networks with binary node states $\sigma_i\in\{0,1\}$ was given by Mc\-Culloch and Pitts \cite{McCulloch:1943} as a model for neural information processing. A model for associative memory constructed from such nodes by Hopfield \cite{Hopfield:1982} attracted considerable interest among physicists as it is conveniently accessible to equilibrium statistical mechanics methods \cite{Amit:1985,Amit:1989,Hertz:1991}. A simple redefinition of weights and thresholds maps the model onto a mathematical representation of a spin glass with states $\sigma_i\in\{-1,1\}$, which has become the usual form of the Hopfield model in the physics literature. The corresponding redefinition of weights and thresholds does not affect the functioning of the model, as its mechanism of  associative memory works on the redefined weights as well. 

In some circumstances, however, when a faithful representation of certain biological details is important, the exact definition matters. In the spin version of a neural network model, for example, a node with negative spin state $\sigma_j=-1$ will transmit non-zero signals through its outgoing weights $c_{ij}$, despite representing an inactive biological node. In the model, such signals arrive at target nodes $i$, e.g., as a sum of incoming signals $h_i= \sum_{j=1}^{N}c_{ij}\sigma_j$. 
However, biological nodes, such as genes or neurons, usually do not transmit signals when inactive. In biochemical network models, each node represents whether a specific chemical component is present $(\sigma=1)$ or absent $(\sigma=0)$. Thus the network itself is mostly in a state of being partially absent, as, e.g., in a protein network where for every absent protein all of its outgoing links are absent as well \cite{Maslov:2002}. In the spin state convention $(\sigma=\pm1)$, this fact is not faithfully represented, whereas networks with Boolean state variables $(\sigma\in{0,1})$ naturally suppress unrealistic signals. 

Another example of an inaccurate detail is the common practice of using the standard convention of the Heaviside step function as an activation function in discrete dynamical networks (or the sign function in a spin model context). The convention $\Theta(0)=1$ is not a careful representation of biological circumstances. For both genes and neurons, a silent input usually maps to a silent output (a gene which is to be active by default is better represented through a threshold value of -1). Therefore, we use a redefined threshold function defined as 
\begin{equation} 
\Theta_0(x)= 
\left\{ 
\begin{array}{c} 
1, \ \ \  x>0 \noindent \\  
0, \ \ \ x \leq 0.  
\end{array} 
\right. 
\end{equation}  

When studying the statistical properties of ensembles of discrete dynamical networks with random links, these details have a considerable influence on the network's dynamics and critical properties. When simulating ensembles via networks of spins, $\sigma_i\in\{-1,1\}$, care should be taken to properly renormalize weights and activation thresholds to ensure a faithful implementation of the original model with states $\sigma_i\in\{0,1\}$. However, this is frequently omitted, e.g.\ in \cite{Kurten:1988,Nakamura:2004,Marro:2008,Kurten:2008,Andrecut:2009}, resulting in the statistics of a system of limited biological plausibility. 

A different field in which normalization and the definition of the nodes' thresholds matters is adaptive networks, currently discussed in the context of neural networks, 
where the critical properties of these networks are of particular interest in the context of self-regulation of activity 
\cite{Bornholdt:2000a,Bornholdt:2003,Arcangelis:2006,Pellegrini:2007,Levina:2007,Gross:2008,Meisel:2009}. 
As with random Boolean networks, threshold networks also exhibit a phase transition from an ordered phase (where limit cycles are typically short and damage does not spread over the system) to a disordered regime (where limit cycle length scales exponentially with system size, and damage may propagate over the entire system). This transition takes place at a critical average connectivity. 
When defining local adaptive mechanisms of self-organized criticality, it is particularly important to base it on biologically plausible definitions of nodes and circuits. While these mechanisms work also for spin type networks \cite{Bornholdt:2003}, such an implementation is not realizable in a biological context, as it would require signals over links which are in fact silent, due to the inactivity of their source nodes. An adaptive algorithm based on such correlations of non-activity is, therefore, not plausible.

In this paper, we first define a binary threshold network that does not include explicitly forbidden states or signals in the context of biological examples. Then, in the main part of the article, we study randomly connected binary threshold networks to derive their critical properties, which we find to be distinctly different from those of random Boolean networks \cite{Kauffman:1969,Derrida:1986,Derrida:1986a,Bastolla:1996,Bastolla:1998} and classical random Boolean threshold networks \cite{Kurten:1988,Kurten:1988a, Rohlf:2002}. In particular, the activity of the network now influences criticality in a non-trivial way, quite similar to random Boolean threshold networks with bistable nodes \cite{Szejka:2008}. In a final section, we will apply the binary threshold network defined here to a specific example from the field of protein networks and gene regulatory networks. We translate two recent cell-cycle network models of budding yeast and fission yeast to the simpler model framework introduced here. 

\section{The Model}
Let us consider threshold networks of $N$ nodes with states $\sigma_i \in \{ 0, 1 \}$, $i \in \{ 1, N \}$, with a random connectivity of on average $K$ links per node. At each discrete time step, all nodes are updated in parallel according to
\begin{equation}
	\sigma_i(t+1) = \Theta_0(f_i(t)) 
	\label{eq:SignumFunction}
\end{equation}
with the redefined threshold function 
\begin{equation}
\Theta_0(0) := 0
\end{equation}
and the input function
\begin{equation}
	f_i(t) = \sum_{j=1}^N c_{ij} \sigma_j(t) + \theta_i.
	\label{eq:InputSum}
\end{equation}
While the weights take discrete values $c_{ij} = \pm 1$ with equal probability for connected nodes, we choose thresholds $\theta_i = 0$ in the following. 
Note that here we strictly consider the dynamics of a discrete dynamical model. 
The justification for such models in its own right, in the context of regulatory network models, has been discussed elsewhere \cite{Bornholdt:2005,Davidich:2008a}.
In certain cases, it is instructive to consider additional noise. Noise can easily be added to our model, for example by a Glauber update rule as we did in our recent paper \cite{Rybarsch:2012}. Please note, however, that then the activation threshold has to be shifted by a value of 0.5, in order to get the transition between the stochastic and the deterministic mode right. In this work, however, we only study the discrete dynamical model where the shift does not have any dynamical consequences and can thus be omitted, such that from now on we set $\theta_i = 0$. 
 
For any node $i$, the number of incoming links $c_{ij} \neq 0$ is called the in-degree $k_i$ of that specific node. With randomly placed links, the probability for each node to actually have $k_i = k$ incoming links follows a Poisson distribution
\begin{equation}
	p(k_i=k) = \frac{{K}^k}{k!} \cdot \exp(-{K}).
	\label{eq:PoissonianLinkDistribution}
\end{equation}

\section{Calculation of the critical connectivity $K_c$}
To analytically derive the critical connectivity $K_c$ of this type of network model, we first study damage spreading on a local basis and calculate the probability $p_s(k)$ for a single node to propagate a small perturbation, i.e. to change its output from 0 to 1 or vice versa after a change of a single input state. The calculation can be done closely following the derivation for spin-type threshold networks in ref.\ \cite{Rohlf:2002}, but one has to account for the possible occurrence of ``0'' input signals also via non-zero links. The combinatorial approach yields a result that directly corresponds to the spin-type network calculation via $p_s^{\mathrm{Bool}}(k) = p_s^{\mathrm{spin}}(2k)$.
\begin{figure}
\epsfig{file=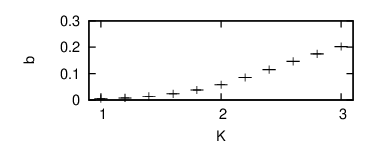, width=8cm}
\caption{Average node activity $b$ as function of connectivity $K$ measured on attractors of 10000 sample networks each, 200 nodes.}
\label{fig:Activity}
\end{figure}
However, this approach does not hold true for our Boolean model in combination with the redefined theta function $\Theta_0(0) := 0$ as it assumes a statistically equal distribution of all possible input configurations for a single node. In the Boolean model, this would involve an average node activity of $b=0.5$ over the whole network (where $b$ denotes the average fraction of nodes which are active, i.e. $\sigma_i=1$). 
Instead we find (Fig.~\ref{fig:Activity}) that the average activity on the network is significantly below $0.5$. At $K=4$ (which will turn out to be already far in the supercritical regime), less than 30 percent of all nodes are active on average. Around $K\approx 2$ (where we usually expect the critical connectivity for such networks), the average activity is in fact below 10 percent. Thus, random input configurations will more likely consist of a higher number of ``0'' signal contributions than of $\pm 1$ inputs.

Therefore, when counting input configurations for the combinatorial derivation of $p_s(k)$, we need to weight all relevant configurations according to their realization probability as given by the average activity $b$. For the first $k = 1,2,3 \ldots$ this yields
\begin{eqnarray*}
	p_s(1) &=& \frac{1}{2} \\
	p_s(2) &=& \frac{1}{2} - b \left( \frac{1}{4} \right) \\
	p_s(3) &=& \frac{1}{2} - b \left( \frac{1}{2} - \frac{1}{4} b \right) \\
	p_s(4) &=& \frac{1}{2} - b \left( \frac{3}{4} - \frac{3}{4} b + \frac{5}{16} b^2 \right) \\
	p_s(5) &=& \ldots \phantom{\left( \frac{1}{2} \right)}
\end{eqnarray*}
which generalizes to
\begin{equation} 
	p_s(k) = \frac{1}{2} + \left( \sum_{i=1}^k (-1)^i \binom{k-1}{i} X_i b^i \right)
	\label{eq:ps_k}
\end{equation}
using $X_i = X_{i-1} \cdot \left( \frac{2i-1}{i+1} \right)$ and $X_1=\frac{1}{4}$.

As the in-degree $k$ is not equal for all nodes, we calculate the average damage propagation per connectivity, which is essential to determine the critical connectivity of the whole network:
\begin{equation}
	\langle p_s \rangle ({K}) = \sum_{k=1}^\infty k \cdot p_k \cdot p_s (k) / C
	\label{eq:<ps_k>}
\end{equation}
where $p_k$ denotes the in-degree distribution from eq. (\ref{eq:PoissonianLinkDistribution}) and $C=\sum_{k=1}^\infty k \cdot p_k = K$ is a normalization constant. We can now compute the sensitivity $\lambda = K \cdot \langle p_s \rangle ({K})$ and apply the annealed approximation to obtain the critical connectivity $K_c$ by solving
\begin{equation}
	\lambda_c = K_c \cdot \langle p_s \rangle (K_c)  = 1. 
	\label{eq:K_cCalculation}
\end{equation}

However, $K_c$ now depends on the average network activity, which in turn is a function of the average connectivity ${K}$ itself, as shown in Fig.~\ref{fig:Activity}.
\begin{figure}
\epsfig{file=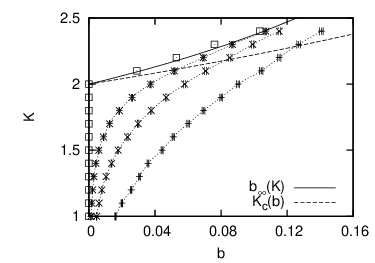, width=8cm}
\caption{Average activity $b$ on attractors of different network sizes (right to left: $N=50, 200, 800$, ensemble averages were taken over 10000 networks each). Squares indicate activity on infinite system determined by finite size scaling, which is in good agreement with the analytic result (solid line). The dashed line shows the analytic result for $K_c(b)$ from eq.\ (\ref{eq:K_cCalculation}). The intersections represent the value of $K_c$ for the given network size.}
\label{fig:ActivityVsKc}
\end{figure}
From the combined plot in Fig.~\ref{fig:ActivityVsKc} we find that both curves intersect at a point where the network dynamics -- due to the current connectivity $K$ -- exhibits an average activity which in turn yields a critical connectivity $K_c$ that exactly matches the given connectivity. This intersection thus corresponds to the critical connectivity of the present network model. 

However, the average activity still varies with different network sizes, which is obvious from Figure~\ref{fig:ActivityVsKc}. Therefore, also the critical connectivity is a function of $N$. Table~\ref{tab:Kc_sizes} lists results for different values of $N$.
\begin{table}
	\begin{center}
		\begin{tabular}{c||c|c|c|c|c|c}
			$N$ & 50 & 100 & 200 & 400 & 800 & $\rightarrow \infty$\\ \hline
			$K_c (\pm 0.005)$ & 2.285 & 2.225 & 2.180 & 2.145 & 2.115 & 2.000
		\end{tabular}
		\caption{Critical Connectivity $K_c$ for different sizes as determined from curve intersections in Figure~\ref{fig:ActivityVsKc}.}
		\label{tab:Kc_sizes}
	\end{center}
\end{table}
For an analytic approach to the infinite size limit, we can now calculate the probability for a node at given in-degree $k$ and average network activity $b_t$ at time $t$ to exhibit output state 1. This probability equals the average activity for the next time step $b_{t+1}$. By examining all relevant input configurations, we find that for a given constant $k$ this generalizes to
\begin{equation} 
	b_{t+1}(k) = \sum_{i=1}^{k} (-1)^{i+1} \frac{1}{2^i} \binom{2i-1}{i-1} \binom{k}{i} b_t^i. 
	\label{eq:b_t+1}
\end{equation}
Again, we have to account for the Poissonian distribution of links in our network model, so the average evolution of network activity is obtained by
\begin{equation}
	\langle b_{t+1} \rangle ({K}) = \exp (-{K}) \sum_{k=1}^\infty \frac{{K}^k}{k!} b_{t+1} (k).
	\label{eq:<b_t+1>}
\end{equation}
It is now possible to distinguish between the different dynamical regimes by solving $\langle b_{t+1} \rangle = b_t(K)$ for the critical line. The solid line in Figure~\ref{fig:ActivityVsKc} depicts the evolved activity in the long time limit. We find that for infinite system size, the critical connectivity is at 
\begin{equation*}
{K}_c(N \rightarrow \infty) = 2.000 \pm 0.001
\end{equation*}
while up to this value all network activity vanishes in the long time limit ($b_\infty=0$). For any average connectivity ${K} > 2$, a certain fraction of nodes remains active. In finite size systems, both network activity evolution and damage propagation probabilities are subject to finite size effects, thus increasing $K_c$ to a higher value.

\begin{figure}
\epsfig{file=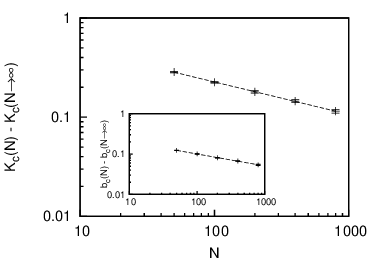, width=8cm}
\caption{Finite size scaling of $K_c(N)$, optimum fit shown here for $K_c(N \rightarrow \infty) = 2.00$. Inset: Finite size scaling of $b(N)$ along the critical line, the optimum fit is obtained for vanishing activity $b_c(N \rightarrow \infty) = 0$.}
\label{fig:Kc-scaling}
\end{figure}

As a numerical verification, we can also derive the critical connectivity for infinite system size $K_c(N \rightarrow \infty)$ using finite size scaling of the above simulation results (Table~\ref{tab:Kc_sizes}). The optimum fit is shown in Figure~\ref{fig:Kc-scaling} and yields
\begin{equation*}
K_c(N \rightarrow \infty) = 2.00 \pm 0.01
\end{equation*}
which supports the analytical calculation.
The same consideration is also possible to obtain the average activity $b(N)$ for increasing network size. This can be done both at constant values of $K$ as well as along the critical line in Figure~\ref{fig:ActivityVsKc} using the values of $K_c(N)$ from Table~\ref{tab:Kc_sizes}. For the critical line, we indeed find vanishing network activity (inset of Figure~\ref{fig:Kc-scaling}):
\begin{equation*}
b_c(N \rightarrow \infty) = 0.00 \pm 0.01.
\end{equation*}
This also holds true for any connectivity below the critical line. For supercritical networks at $K > K_c$, the numerical simulation yields a non-zero fraction of nodes which remains active on average, in good agreement with the analytic computation (see squares in Figure~\ref{fig:ActivityVsKc}).

In additional numerical simulations using the standard step function, we obtain a critical connectivity of $K_c \approx 3.7$, which is analytically supported by a combinatorial calculation following ref.\ \cite{Rohlf:2002}, where we find $p_s^{\mathrm{Bool}}(k) = p_s^{\mathrm{spin}}(2k)$. The networks exhibit significantly higher average activity, while most of the active nodes are frozen in the active state. On a side note, if we chose to calculate the critical connectivity based on an assumed average activity of $0.5$, the activity-dependent calculation via (\ref{eq:ps_k}) and  (\ref{eq:K_cCalculation}) would effectively reproduce the same result ($K_c \approx 3.7$) as the original combinatiorial approach from \cite{Rohlf:2002} would yield for the new Boolean model. As the assumption does not hold true here, this result can only be viewed as an additional plausibility check for the correspondence between both approaches.

\begin{figure}
\epsfig{file=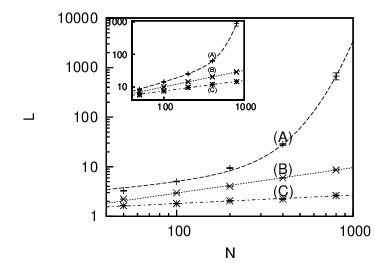, width=8cm}
\caption{Average attractor length at different network sizes. Ensemble averages were taken over 10000 networks each at (A) $K=2.4$, (B) $K=2.0$, (C) $K=1.6$. Inset figure shows the scaling behavior of the corresponding transient lengths.}
\label{fig:Attractors}
\end{figure}
Finally, let us take a closer look at the average length of attractor cycles and transients. As shown in Fig.~\ref{fig:Attractors}, the behavior is strongly dependent on the dynamical regime of the network. As expected and in accordance with early works on random threshold networks \cite{Kurten:1988} as well as random Boolean networks \cite{Bastolla:1996}, we find an exponential increase of the average attractor lengths with network size $N$ in the chaotic regime ($K>K_c$), whereas we can observe a power-law increase in the frozen phase ($K<K_c$). We find similar behavior for the scaling of transient lengths (inset of Figure~\ref{fig:Attractors}).

After this characterization of the dynamical properties and critical behavior of random networks of the proposed node type, we will now demonstrate that it is also suited for a particularly simple implementation of discrete dynamical models of biological regulatory networks. 

\section{Binary threshold network model of the yeast cell-cycle control}
Two recent examples of biological networks that can be modeled with Boolean networks are the cell-cycle networks of budding yeast \cite{Li:2004} and fission yeast \cite{Davidich:2008}. In both yeasts, the control of cell division is achieved by a small network of about a dozen genes and associated proteins each. In both cases, a Boolean network model of this network generates a temporal activation pattern that accurately mimics the succession of protein activities in the living cells along cell division. The node type used in these models so far is a relatively complicated bistable threshold node, with extra self-inhibitory loops needed for degradation of proteins \cite{Li:2004}. With the node type studied in this article, these models no longer need bistable nodes. We instead update the node states according to 
\begin{equation}
\sigma_i(t+1)= 
\left\{ 
\begin{array}{c} 
1, \ \ \  \sum_j c_{ij} \sigma_j(t)  >0 \noindent \\  \\
0, \ \ \  \sum_j c_{ij} \sigma_j(t)  \leq 0 \noindent
\end{array} 
\right. 
\end{equation}  
in accordance with the definitions proposed above.
\begin{figure}
\epsfig{file=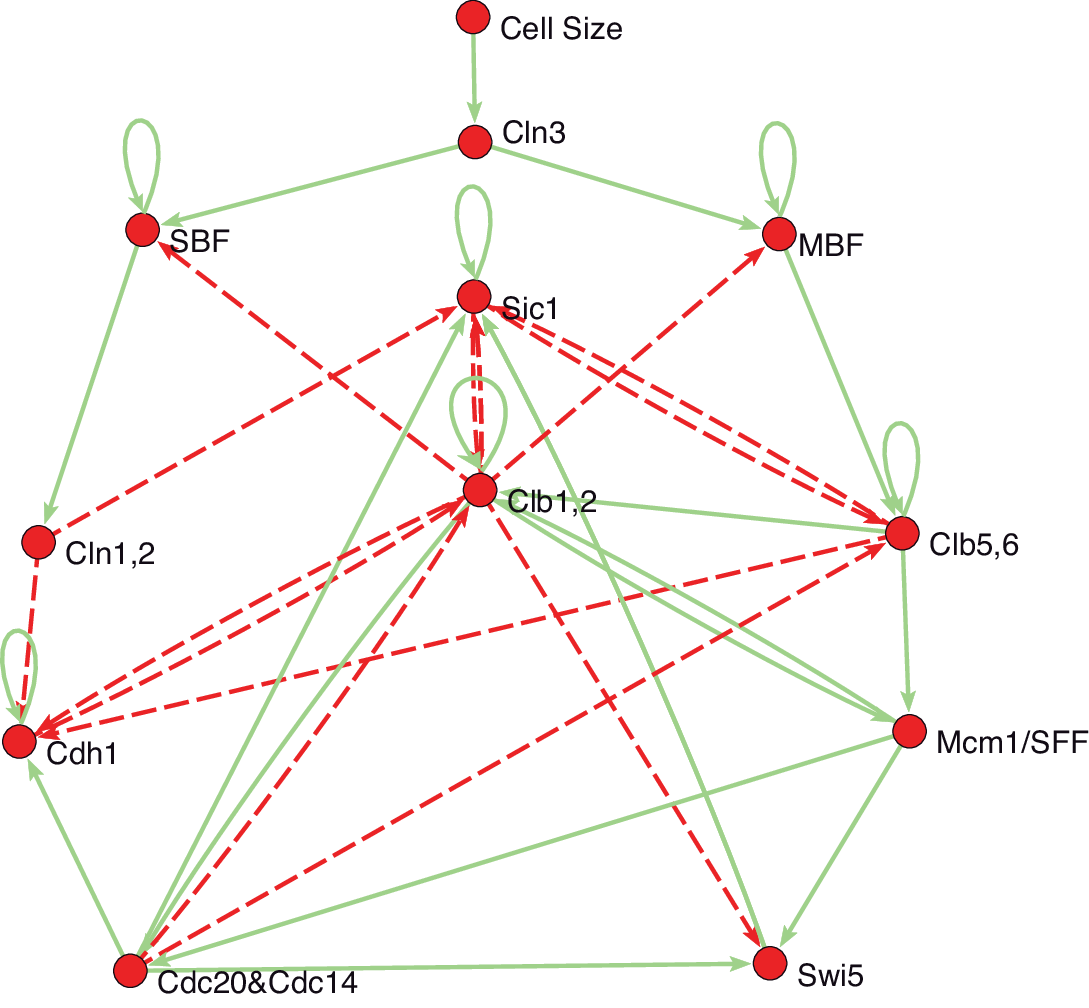, width=8cm}
\epsfig{file=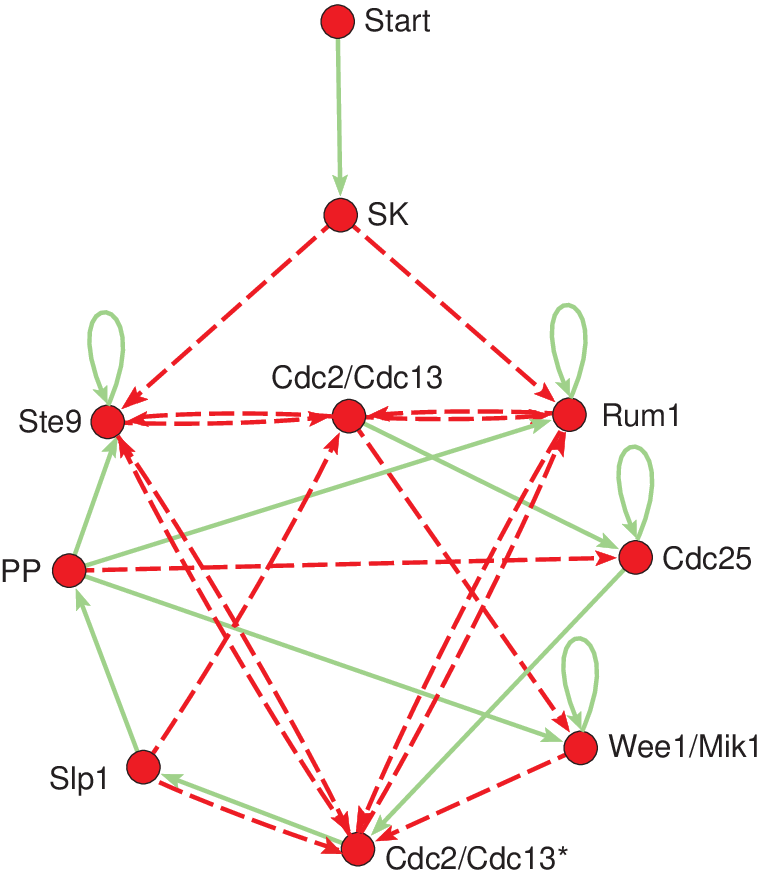, width=6cm}
\caption{(Color online) Practical realization of gene regulatory network models which predict the cell-cycle sequences of budding yeast (top) and fission yeast (bottom). The models from \cite{Li:2004} and \cite{Davidich:2008} are shown as transferred to the present network model, thereby lifting the need for bistable nodes and additional self-degradation. Solid green (light gray) lines depict activating links ($c_{ij}=+1$); dashed red (dark gray) links are inhibitory ($c_{ij}=-1$).}
\label{fig:Yeast}
\end{figure}
The resulting network model with intuitively defined threshold dynamics without bistability is not only constructed more simply, but it is also more plausible in the context of biochemical networks. 
A gene, that is not bistable or active by default, no longer has to 
have a topological element as a self-inhibiting loop (as was done, e.g., in the original model by Li et al. 
as well as many subsequent papers) just to account for a non-topological
dynamical feature such as the trivial degradation of proteins in the cell.  
The transformation to the new threshold nodes from the original models is straightforward: nodes which have a self-degradation self-link in the original models \cite{Li:2004,Davidich:2008} will lose the inhibitory loop in our model. The other nodes, which do not self-degrade in \cite{Li:2004,Davidich:2008} due to bistability, will instead get a self-activation link in our model (except for CellSize/Start nodes). Figure~\ref{fig:Yeast} shows the resulting networks. In both networks, all nodes have zero activation thresholds, $\theta_i = 0$, except for the node \mbox{Cdc2/Cdc13}, which has a threshold value of $\theta_i = -1$. 
\begin{figure}
\epsfig{file=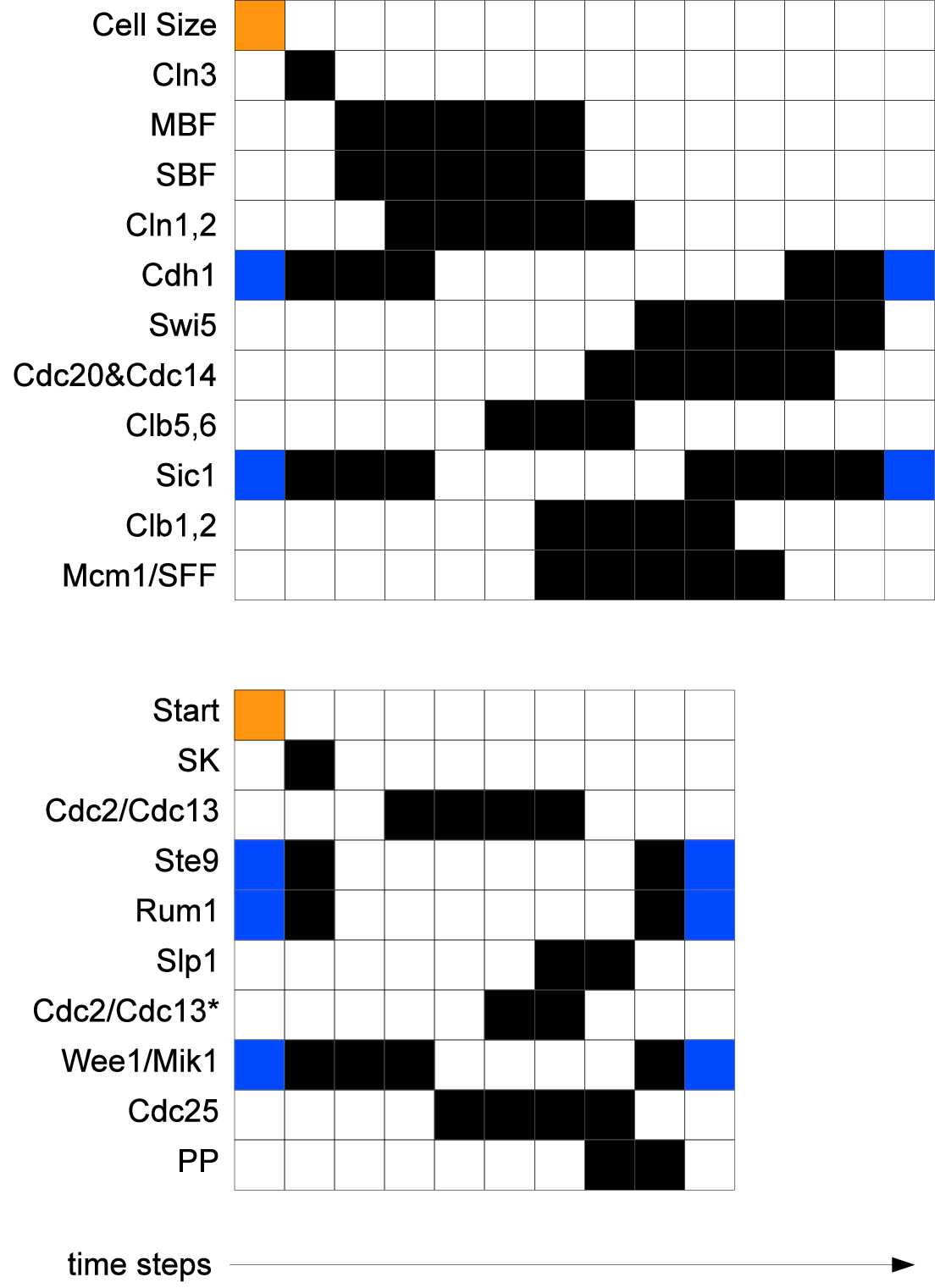, width=8cm}
\caption{(Color online) Temporal evolution of activation states in the cell-cycles of budding yeast (top) and fission yeast (bottom). Results in our model are the same as for the wild-type sequence in \cite{Li:2004,Davidich:2008}. Blue (dark gray) squares depict the activated nodes in the G1 stationary states (fixed points), orange (light gray) squares represent the initiation of the cell-cycle sequence by an external signal when the cell has sufficiently grown.}
\label{fig:Yeast-States}
\end{figure}
When simulated by parallel synchronous updates, both model networks reproduce the correct temporal cell-cycle activation pattern sequence of the respective yeast cells (Figure~\ref{fig:Yeast-States}). While it is obvious that the bistable node behavior used by the older models \cite{Li:2004,Davidich:2008} cannot necessarily be mapped in any case onto our model, it is interesting to note that for these two seminal examples, a transformation to a more simply defined network model without bistable nodes is easily possible. At the same time, self-degradation does not need to be explicitly modeled into the network via self-inhibitory links anymore, but is now an intrinsic feature of the network. The inclusion of self-activating loops in our model is a more plausible representation, as the gene (or protein) which causes the self-activation is actually active (or present) in the network.

\section{Conclusions}
In summary we studied threshold networks with Boolean node states that are biologically more plausible than current Boolean and threshold networks and which are simpler than the recently introduced networks with bistable threshold nodes \cite{Li:2004,Davidich:2008}. A major observation is that the activity of the nodes depends on connectivity, which also renders critical properties of the networks activity-dependent, as found earlier for random threshold networks with bistable nodes \cite{Szejka:2008}. We extend the annealed approximation to correct for these effects and find connectivity $K_c=2.0$ and vanishing activity at the critical point in the thermodynamic limit. 

Going beyond the statistics of random network ensembles, also real biological circuits can be implemented with great ease using the threshold networks defined here. We successfully reproduced the dynamical trajectory of the budding yeast cell cycle network as implemented with bistable threshold functions in \cite{Li:2004}, as well as for the corresponding network in fission yeast \cite{Davidich:2008}, with only minor changes in the network. 

To conclude, let us remind ourselves of the original idea of using random Boolean networks for characterizing typical properties of biological networks \cite{Kauffman:1969}. In 1969, using random Boolean networks as a null model for genetic networks was a logical approach, given our complete ignorance of the circuitry of genetic networks at that time. Thus the best guess was to treat all possible Boolean rules as equally probable. Today, however, we have much more detailed knowledge about certain properties of genetic networks and, therefore, about more realistic ensembles of random networks. A biologically motivated and carefully defined threshold network, as attempted in this paper, may provide a more suited null model for the particular properties of biological networks than random Boolean networks with equally distributed Boolean functions. 

Moreover, our model is not only suitable to gene regulatory networks, as shown above, but it is also plausible as a minimal model for neural networks. While retaining the stunning simplicity of older spin models for self-organized critical neural networks, the present work contributes to a more realistic modeling of neuronal activity characteristics, rendering ensemble statistics with more plausible dynamic properties, e.g., states of low neuronal activity with the occurrence of power-law distributed activity avalanches. Also, the implementation of a biologically inspired mechanism for self-organization being robust against external perturbation is quite possible in the network model presented herein \cite{Rybarsch:2012}.

\section{Acknowledgements}
We acknowledge the support of the DFG under contract
No. INST 144/242-1 FUGG. M.\ Rybarsch would like to thank 
the Studienstiftung des deutschen Volkes for financial support of this work.

\end{document}